\documentstyle[12pt]{article}
\newlength{\cmd}
\topmargin - 10 mm

\begin{document}

\vspace{10  mm}

\begin{center}
{\large {\bf Option Pricing Model for Incomplete Market}}

{\large {\bf \vspace{10  mm} Sergei Fedotov \vspace{5  mm } }}

{\large  Department of Mathematics \\ UMIST PO Box 88 \\ 
Manchester M60 1QD \\ United Kingdom \vspace{2  mm } }

{\large {\bf \vspace{10  mm} Sergei Mihkailov \vspace{5  mm } }}

{\large Bergische Universit\"at \\FB-14, Gaussstr. 20, 42097 Wuppertal \\%
Deutschland}
\end{center}

\vspace{25  mm}

\begin{center}
{\large {\bf ABSTRACT}}
\end{center}

\vspace{10  mm}

\baselineskip \cmd

The problem of determining the European-style option price in the incomplete
market has been examined within the framework of stochastic optimization. An
analytic method based on the discrete dynamic programming equation (Bellman
equation) has been developed that gives the general formalism for
determining the option price and the optimal trading strategy (optimal
control policy) that reduces total risk inherent in writing the option.
The basic purpose of the paper is to present an effective algorithm that
can be used in practice.


\newpage

\section{Introduction}

An essential feature of the currently dominant
option pricing theory proposed by Black and Scholes is the existence of a
dynamic trading strategy in the underlying asset that exactly replicates the
derivative contract payoff [1-4]. However, in general, the market is not
complete, the contingent claim is not redundant asset and therefore its
price cannot be determined by the no-arbitrage argument alone. The reasons
that give rise to an incompleteness of market might be very different, for
example, mixed jump-diffusion price process for an asset [1,5], stochastic
volatility [6], etc.

In recent years there has been a substantial theoretical efforts to give the
pricing formula for a derivative security for which an exact replicating
portfolio in the underlying asset ceases to exist. The typical example
involving incompleteness is a model in which the stock volatility is a
stochastic process. Several approaches to the valuation of the contingent
claim under random volatility have been suggested in literature [6-11].
Typically the pricing formulas involve the unknown and what is more
unobservable parameter, so-called market price of volatility risk. This
fundamental difficulty has led the researches to accept the idea of
uncertain volatility when all prices for contingent claim are possible
within some specified range [12-15].

An alternative method for the derivative pricing in the incomplete market
has been proposed in a series of papers by mathematicians M\"uller,
F\"olmer, Sondermann, Schweizer and Sch\"al [16-20] and by physicists
Bouchaud and Sornette [21] (see also [22-25]).
The basic idea is that the fair price of
contingent claim can be found through the risk minimization procedure.
Different criteria for a measuring the risk inherent in writing an option
have been suggested including the global and local variance of the cost
process [16-18] and the variance of the global operator wealth [19,21-23]. We
refer to the recent survey paper [20] for an exposition of the status of
research on the incomplete market involving the stochastic volatility and
risk-minimization.

Although significant progress has already been made in the option pricing
theory involving the risk minimization procedure, still there exist many
open problems including how to derive an {\it effective algorithm} giving
the option price and trading strategy. The purpose of this paper is to
present such an algorithm that can be used in practice. The aim is to show
how the problem of option pricing based on the risk-minimization analysis
can be reformulated in terms of Maier problem and how the stochastic
optimization procedure [26] based on the Bellman equation can be implemented
to give a reliable numerical technique for determining both the derivative
price and optimal trading strategy. The application of dynamic programming
approach to option pricing can be found in [18,28,29].

\section{Statement of the problem}

We assume a discrete-time $N$-period world in which the dynamics of the
security price $S_n$ is governed by the stochastic difference equation 
$$
S_{n+1}=S_n+\xi _nS_n,\hspace{.3in}n=0,1,...,N-1,\eqno(1) 
$$
where $\xi _n$ is a sequence of random variables whose conditional jump
density at time $n$ is independent of the asset price $S_m$ , $m<n$, and is
given by 
$$
\rho _n\left( \xi ,S_n\right) \equiv \frac \partial {\partial \xi }P\left\{
\xi _n<\xi \mid _{S_n}\right\}. \eqno(2) 
$$

We assume that at time $0$ an investor sells an European-style option with
the strike price $X$ for $C_0$ and invests this money in a portfolio
containing $\Delta _0$ shares held long partially financed by borrowing $B_0$
in cash. The current value of this portfolio is given by 
$$
V_0=C_0=\Delta _0S_0-B_0.\eqno(3) 
$$

The investor is interested in constructing the self-financing strategy to
hedge the option exposure. Since for the incomplete market the exact
replication of the option payouts by a portfolio of traded securities is not
possible, the investor cannot completely neutralize the risk inherent in
writing the option. Hence the problem is to find such a trading strategy
that reduces total risk to some intrinsic value[16-25].

To proceed further we need an equation governing the dynamics of the
self-financing hedged portfolio. First we consider the case of frictionless
trading (the effect of transaction costs will be also examined in this
paper). The value of the portfolio $V_n$ at time $n$ may be written as 
$$
V_n=\Delta _nS_n-B_n,\hspace{.3in}n=0,1,...,N-1,\eqno(4) 
$$
where $\Delta _n$ is the number of shares of the underlying asset held long
during the time interval $[n,n+1)$ and $B_n$ is the amount of money
borrowed. At the beginning of trading period $n+1$ just before readjusting
the position this portfolio is worth 
$$
V_{n+1}=\Delta _nS_{n+1}-\left( 1+r\right) B_n,\hspace{.3in}n=0,1,...,N-1,%
\eqno(5) 
$$
where $r$ is the interest rate. Therefore the change in the value of the
portfolio can be written as 
$$
V_{n+1}=\left( 1+r\right) V_n+\Delta _n\left( \xi _n-r\right) S_n,%
\hspace{.3in}n=0,1,...,N-1.\eqno(6) 
$$

Following [19] we propose that the investor's purpose is to maintain a
self-financing portfolio (4) in a such way that at the expiration date $N$
the terminal value of this portfolio 
$$
V_N=\left( 1+r\right) V_{N-1}+\Delta _{N-1}\left( \xi _{N-1}-r\right) S_{N-1}%
\eqno(7) 
$$
should be as close as possible to the option payoff 
$$
\theta _X\left( S_N\right) \equiv \max \left( S_N-X,0\right) .\eqno(8) 
$$

One way to achieve this purpose is to require that the expectation value of
the difference between the option value and the value of hedged portfolio at
expiration is equal to zero, i.e. ${\bf E}\left\{ \theta _X\left( S_N\right)
-V_N\right\} =0$ while the variance of this difference ${\bf E}\left\{
\left( \theta _X\left( S_N\right) -V_N\right) ^2\right\} $ as a measure for
risk should be minimized by the proper choice of the trading strategy; here $%
{\bf E}\left\{ \cdot \right\} $ denotes expectation with respect to the
distributions of $\xi _0,$ $\xi _1,...,\xi _{N-1}.$

\section{Stochastic optimization}

According the ideas of dynamic programming [26,27], the proper choice of the
sequence controls $\Delta _n,B_n$ should involve the information
aggregation, i.e. the optimal choice of trading strategy at each of $N$ time
period should be based on the available information about the current values
of asset price and hedged portfolio. From a mathematical point of view it
means that one have to find a sequence of functions (so-called optimal
control policy) 
$$
\Delta _n^{*}=\Delta _n^{*}\left( S_n,V_n\right) ,\hspace{.2in}%
B_n^{*}=B_n^{*}\left( S_n,V_n\right) \hspace{.3in}n=0,1,...,N-1\eqno(9) 
$$
that minimize the total risk. In what follows we will use (4) to find an
optimal value of $B_n$, that is. 
$$
B_n^{*}\left( S_n,V_n\right) =\Delta _n^{*}\left( S_n,V_n\right) S_n-V_n,%
\hspace{.3in}n=0,1,...,N-1.\eqno(10) 
$$

Instead of the problem of minimizing the risk subject to the constraint we
consider the problem of minimizing the modified risk-function 
$$
R_\lambda \equiv {\bf E}\left\{ \left( \theta _X\left( S_N\right)
-V_N\right) ^2+\lambda \left( \theta _X\left( S_N\right) -V_N\right)
\right\} ,\eqno(11) 
$$
where $\lambda $ is the Lagrange multiplier and the average is made again
over all $\xi _n$.

Following the dynamic programming approach, we consider first the last time
period and proceed backward in time. If at the beginning of the last trading
period $N-1$ the stock price is $S_{N-1}$ and the value of portfolio is $%
V_{N-1}$ , then no matter what happened in the past periods, the investor
should choose such a trading strategy $\Delta _{N-1},B_{N-1}$ that minimizes
the risk for the last time period.

Let us introduce the minimal risk for the last period which is a function of
the stock price $S_{N-1}$ and the value of the portfolio $V_{N-1}$ 
$$
I_0\left( S_{N-1},V_{N-1}\right) =\min_{\Delta _{N-1}}{\bf E}_{\xi
_{N-1}}\left\{ \left( \theta _X\left( S_N\right) -V_N\right) ^2+\lambda
\left( \theta _X\left( S_N\right) -V_N\right) \right\} .\eqno(12) 
$$
It follows from (1) and (6) that $I_0$ can be rewritten as 
$$
I_0\left( S_{N-1},V_{N-1}\right) =
$$
$$
\min_{\Delta _{N-1}}{\bf E}_{\xi
_{N-1}}\left( \theta _X\left( S_{N-1}+\xi _{N-1}S_{N-1}\right) -\left(
1+r\right) V_{N-1}-\Delta _{N-1}\left( \xi _{N-1}-r\right) S_{N-1}\right)^2 
$$
$$
+\lambda \left( \theta _X\left( S_{N-1}+\xi _{N-1}S_{N-1}\right) -\left(
1+r\right) V_{N-1}-\Delta _{N-1}\left( \xi _{N-1}-r\right) S_{N-1}\right) .%
\eqno(13) 
$$
By calculating this function we obtain the optimal value of $\Delta _{N-1}$
and thereby the optimal trading policy $\Delta _{N-1}^{*}\left(
S_{N-1},V_{N-1}\right) ,B_{N-1}^{*}\left( S_{N-1},V_{N-1}\right) $ for the
last period.

At the beginning of time period $N-2$ when the stock price is $S_{N-2}$ and
the value of portfolio is $V_{N-2}$ the investor should readjust the
position in a such way that $\left( \Delta _{N-2},B_{N-2}\right) $ minimize
the risk ${\bf E}_{\xi _{N-2}}\left\{ I_0\left( S_{N-1},V_{N-1}\right)
\right\} $ .

The dynamic programming algorithm takes the form of the recurrence relation 
$$
I_1\left( S_{N-2},V_{N-2}\right) =
$$
$$
\min_{\Delta _{N-2}}{\bf E}_{\xi
_{N-2}}\left\{ I_0\left( S_{N-2}+\xi _{N-2}S_{N-2},\left( 1+r\right)
V_{N-2}+\Delta _{N-2}\left( \xi _{N-2}-r\right) S_{N-2}\right) \right\} .%
\eqno(14) 
$$
By calculating $I_1\left( S_{N-2},V_{N-2}\right) $ we obtain the optimal
function $\Delta _{N-2}^{*}=$ $\Delta _{N-2}^{*}\left(
S_{N-2},V_{N-2}\right) $ .

Repeating these arguments we can get the Bellman equation for the period $n$ 
$$
I_{N-n}\left( S_n,V_n\right) =\min_{\Delta _n}{\bf E}_{\xi _n}\left\{
I_{N-n-1}\left( S_n+\xi _nS_n,\left( 1+r\right) V_n+\Delta _n\left( \xi
_n-r\right) S_n\right) \right\} .\eqno(15) 
$$

The last equation can be rewritten in the form 
$$
I_{N-n}\left( S_n,V_n\right) =\min_{\Delta _n}\int I_{N-n-1}\left( S_n+\xi
S_n,\left( 1+r\right) V_n+\Delta _n\left( \xi -r\right) S_n\right) \rho
_n\left( \xi ,S_n\right) d\xi .\eqno(16) 
$$

The attractive feature of the dynamic programming algorithm is the relative
simplicity with which the optimal trading policy $\Delta _n^{*}=\Delta
_n^{*}\left( S_n,V_n\right) ,B_n^{*}=B_n^{*}\left( S_n,V_n\right) $ , $%
n=0,1,...,N-1$ can be computed. The basic advantage of general algorithm
(15) over functional derivative technique [21] is that the original problem
(11) is reduced to a sequence of minimization problems which of them is much
simpler than the original one.

It might seem that the better choice of control in (15) would be a pair $%
\left( \Delta _n,B_n\right) $ giving the control policy $\Delta _n^{*}\left(
S_n\right) ,B_n^{*}\left( S_n\right) $ as the functions of the asset price $%
S_n$ only. However the self-financing condition gives rise to the
restriction 
$$
\Delta _nS_n-B_n=\Delta _{n-1}S_n-\left( 1+r\right) B_{n-1}\eqno(17) 
$$
which makes the control problem in terms of the pair $\left( \Delta
_n,B_n\right) $ rather difficult.

It is clear that the function $I_N\left( S_0,V_0\right) $ is the minimal
risk for the optimal trading strategy when the initial value of stock is $%
S_0 $ and the value of portfolio is $V_0.$ The initial investment required
to fund the partially hedged portfolio is nothing else but the price of
option $C_0$ which can be determined from the equation 
$$
\frac{\partial I_N\left( S_0,C_0\right) }{\partial C_0}=0.\eqno(18) 
$$

The discussion of when the optimal initial investment $C_0$ can be
considered as a fair option price and problems that might arise from that
can be found in [19].

\section{Transactions costs}

Let us now consider the problem of finding the optimal trading strategy and
the option price in the presence of transactions costs. We know that the
effects of transactions costs on the contingent claim pricing might be very
complex depending on the size of bid-offer spreads, the structure of payoff
functions, etc.[4,30-32]. Here we suggest a new algorithm for a valuation of
option price based on the risk minimization procedure.

We assume a bid-offer spread in which the investor buys the stock for the
offer price $S\left( 1+k\right) $ and sells it for the bid price $S\left(
1-k\right) $. Again we formulate the problem in terms of an investor who
sells the European option with payout $\theta _X\left( S_N\right) $ and who
employs the trading strategy to hedge the derivative. At time $0$ a hedged
portfolio is constructed by purchasing of $\Delta _0$ shares at the offer
price $S_0\left( 1+k\right) $ and borrowing $B_0$ in cash at the riskless
rate $1+r$ , so that the ammount of money spent for this portfolio including
the effect of transaction cost can be written as 
$$
V_0=\Delta _0S_0-B_0+k\Delta _0S_0.\eqno(19)
$$
It is assumed here that the investor has no initial position in the
underlying asset. The investor's purpose is to maintain a dynamic portfolio
strategy in a such way that the risk of his liability (11) is minimal.

Above we have derived a Bellman equation (15) when the asset price and the
value of portfolio have been chosen as the dynamical variables while the
number of shares in portfolio has played the role of the control parameter.
In the presence of transaction costs it is more convenient to make another
choice of basic variables and controls. It follows from the self-financing
condition that the stochastic dynamics of the amount of dollars $B_n$
borrowed can be written as 
$$
B_{n+1}=\left( 1+r\right) B_n+\left( \Delta _{n+1}-\Delta _n+k\mid \Delta
_{n+1}-\Delta _n\mid \right) S_{n+1},\hspace{.3in}n=0,1,...,N-1.\eqno(20) 
$$

Let us introduce a new control parameter $\Omega _n$ such that 
$$
\Delta _{n+1}=\Omega _n,\hspace{.3in}n=0,1,...,N-1,\eqno(21)
$$
then the dynamical variables $B_n$ obeys the stochastic difference equation 
$$
B_{n+1}=\left( 1+r\right) B_n+\left( \Omega _n-\Delta _n+k\mid \Omega
_n-\Delta _n\mid \right) \left( 1+\xi _n\right) S_n,\hspace{.3in}%
n=0,1,...,N-1.\eqno(22)
$$
Now we are in a position to formulate the basic problem. Let us denote by $%
I_N\left( S_0,B_0,\Delta _0\right) $ the minimal risk that can be achieved
by starting from the arbitrary initial state $S_0,B_0,\Delta _0$ 
$$
I_N\left( S_0,B_0,\Delta _0\right) =\min_{\Omega _0,...,\Omega _{N-1}}{\bf E}%
\left\{ \left( \theta _X\left( S_N\right) -\Delta _NS_N+B_N\right)
^2+\lambda \left( \theta _X\left( S_N\right) -\Delta _NS_N+B_N\right)
\right\} .\eqno(22)
$$

Then the principle of optimality yields the general recurrence relation 
$$
I_{N-n}\left( S_n,B_n,\Delta _n\right) =
$$
$$
\min_{\Omega _n}{\bf E}_{\xi _n}\left\{ I_{N-n-1}\left( S_n+\xi _nS_n,\left(
1+r\right) B_n+\left( \Omega _n-\Delta _n+k\mid \Omega _n-\Delta _n\mid
\right) \left( 1+\xi _n\right) S_n,\Omega _n\right) \right\} .\eqno(23)
$$

Clearly, the optimal values of $\Delta _0$ and $B_0$ determining the initial
investment $V_0$ as a fair option price can be determined by 
$$
\frac{\partial I_N\left( S_0,B_0,\Delta _0\right) }{\partial B_0}=0,%
\hspace{.3in}\frac{\partial I_N\left( S_0,B_0,\Delta _0\right) }{\partial
\Delta _0}=0.\eqno(24)
$$


\section{Summary}

To conclude, we have formulated the European-style pricing model for the
incomplete market in which the risk incurred by selling an option cannot be
completely hedged by dynamic trading. New effective algorithm based on the
discrete dynamic programming equation has been presented that gives the
option price and optimal trading strategy. The method accommodates the
effects of transaction costs and can easily be extended to price
options when volatility is random. The uncertain volatility case can be also
treated by the stochastic optimization procedure.
There are several directions to explore by the method presented here.
First, one may study the case with imperfect state information regarding
the asset prices. Also, one can study various adaptive control problems. 
It should be noted that the preliminary work done here might be of big
practical importance and therefore merits further investigation including
he computational aspect of our formalism.

\end{document}